# Deuterium fractionation on interstellar grains studied with modified rate equations and a Monte Carlo approach

Paola Caselli[a], Tatiana Stantcheva[b], Osama Shalabiea,[c] Valery I. Shematovich[d] and Eric Herbst[e]

[a] Osservatorio Astrofisico de Arcetri, Largo E. Fermi 5, I-50125 Firenze, Italy
[b] Department of Physics, The Ohio State University, Columbus, OH 43210 USA
[c] Department of Astronomy, Faculty of Science, Cairo University, Egypt
[d] Institute of Astronomy of the Russian Academy of Sciences, 48 Pyatnitskaya St., Moscow 109017, Russia
[e] Departments of Physics and Astronomy, The Ohio State University, Columbus, OH 43210 USA

**Abstract**

The formation of singly and doubly deuterated isotopomers of formaldehyde and of singly, doubly, and multiply deuterated isotopomers of methanol on interstellar grain surfaces has been studied with a semi-empirical modified rate approach and a Monte Carlo method in the temperature range 10-20 K. Agreement between the results of the two methods is satisfactory for all major and many minor species throughout this range. If gas-phase fractionation can produce a high abundance of atomic deuterium, which then accretes onto grain surfaces, diffusive surface chemistry can produce large abundances of deuterated species, especially at low temperatures and high gas densities. Warming temperatures will then permit these surface species to evaporate into the gas, where they will remain abundant for a considerable period. We calculate that the doubly deuterated molecules  CHD$_2$OH and CH$_2$DOD are particularly abundant and should be searched for in the gas phase of protostellar sources. For example, at 10 K and high density, these species can achieve up to 10-20% of the abundance of methanol.

## 1. Introduction

High abundances of doubly deuterated molecular isotopomers have been detected in several sources in interstellar clouds. Although the abundance of $ND_2H$ in the cold clump L134N can probably be accounted for by gas-phase fractionation (Roueff et al., 2000; Roberts and Millar, 2000; Rogers and Charnley, 2001) with significant depletion of heavy elements and molecules composed of them, the abundance of $D_2CO$ in warm star-forming regions cannot be so explained. Originally detected by Turner (1990) in the Orion compact ridge with an abundance $fD_2CO \equiv D_2CO/H_2CO \approx 0.003$, this species has now been detected towards IRAS 16293-2422 with $fD_2CO \approx 0.1$ (Ceccarelli et al., 1998, 2001; Loinard et al., 2000). It is most likely that the high abundance of $D_2CO$ in warm objects derives from the occurrence of grain-surface chemistry during an earlier cold era followed by evaporation into the gas as temperatures rise (Ceccarelli et al., 2001; Charnley et al., 1997). In this view, gas-phase processes operating at low temperatures first produce a high abundance of atomic D compared with atomic H (Tielens, 1983; Brown and Millar, 1989). These atoms, upon accreting onto the surfaces of dust particles, then convert a portion of the surface CO into the hydrogenated species $H_2CO$ and $CH_3OH$ and their singly, doubly, and even (for the case of methanol) triply and quadruply deuterated isotopomers (Tielens, 1983; Charnley et al., 1997). The conversion requires reactions with at least some activation energy and, depending upon the rate of diffusion of H and D on grain surfaces, may take a considerable amount of time.

A first attempt at modeling the surface chemistry that converts CO into formaldehyde and methanol and their deuterated isotopomers was published by Charnley et al. (1997). They considered a cold high-density cloud in which the hydrogen is mainly converted into molecular hydrogen and the carbon into CO. Gas-phase chemical models of dense clouds show that such a situation takes at least $10^5$ yr to develop (Terzieva and Herbst, 1998). Residual atomic hydrogen, deuterium, and CO were allowed to accrete onto grain mantles and produce formaldehyde, methanol, and their isotopomers via association reactions between H and D atoms and unsaturated molecules. For the high densities considered, the accretion rate of H atoms is smaller than that of CO molecules.

Charnley et al. (1997) used a probabilistic method (see also Brown and Millar, 1989) to determine relative abundances of these species with respect to each other at times sufficiently long that steady-state is reached. In this condition, the surface abundances of all major species increase at the same rate so that their relative abundances remain the same. The surface chemistry results of Charnley et al. are shown in their Figures 2 and 3. In their Figure 2, the relative abundances of CO and $H_2CO$ to $CH_3OH$ are plotted against one another for differing relative accretion rates of H and CO and for differing relative activation energy barriers between the two reactions – H + CO and H + $H_2CO$. In their Figure 3, the ratios $f$ of $D_2CO$ and assorted singly deuterated isotopomers to their non-deuterated counterparts are plotted vs. the relative grain accretion rate of D to CO under conditions thought to be relevant for the Orion Hot Core/compact ridge. The $f$ values can become quite high for large accretion rates of D atoms; the analysis can possibly explain the observations of $D_2CO$ in warm regions if gas-phase fractionation can build up a rather large atomic D/H ratio in the gas.

The analysis of Charnley et al. (1997) contains a strong assumption that allows simplification of the mathematics: the H and D atoms both migrate so rapidly on grains that they react rather than evaporate even though the initial reaction with CO has an activation energy barrier! Charnley et al. made this assumption based on then current ideas concerning the diffusion rate of H. Recent experimental work by Vidali, Pirronello, and co-workers (see, e.g. Katz et al., 1999) indicates that H atoms appear to move much more slowly on two realistic analogs of grain surfaces – olivine and amorphous carbon. Evaporation of H and D atoms prior to reaction is then important even at 10 K until a large build-up of surface CO can occur. At higher temperatures, evaporation can be important even with rapid diffusion rates.

In this paper, we report the results of a somewhat more comprehensive model of the surface chemistry of formaldehyde and methanol and their deuterated isotopomers than that of Charnley et al. (1997). In our model, we include the surface formation of molecular hydrogen, carbon dioxide, and water ice as well as the deuterated species HD, $D_2$, HDO, and $D_2O$. We consider a variety of diffusion rates, for some of which the evaporation of the H and D reactants can be important, and we vary a variety of parameters (temperature, density, activation energy barriers). The time needed to build up observable mantle populations of the species is explicitly considered. We are especially interested in the calculated abundances of doubly-deuterated isotopomers of methanol; these species have not yet been detected in star-forming regions but are exciting candidates for detection. Another major purpose of this paper is to test the semi-empirical modified rate method of modeling surface chemistry (Caselli et al., 1998; Stantcheva et al., 2001) with a more detailed Monte Carlo method. We have previously used the modified rate method in large-scale chemical models of the chemistry occurring in the gas and on the grains in dense clouds (Ruffle and Herbst 2000, 2001a, 2001b) but have so far only been able to test it with a non-empirical pseudo-Monte Carlo approach for the extremely simple O, H system (Stantcheva et al., 2001). In this latter system, gaseous O and H atoms accrete onto grain surfaces and react to form the diatomic molecules $O_2$, $H_2$, and OH. We show here that the modified rate method is a reasonable if not perfect approximation to the Monte Carlo kinetics for the deuterated methanol system throughout much of the parameter range explored so far, which is representative of translucent and dense clouds. The Monte Carlo method used is itself different from our previous approaches and is based on the chemical master equation (Gillespie, 1976; Charnley, 1998).

In a future paper, we will look in more detail at this Monte Carlo approach and compare its results for the methanol system with a direct solution of the master equation (Biham et al., 2001; Green et al., 2001).

## 2. Model

As did Charnley et al. (1997), we consider a cold interstellar cloud that has evolved to the stage where most of the hydrogen is molecular and much of the gas-phase carbon is in the form of CO. On the grains, appreciable amounts of water ice have already been formed by hydrogenation of accreting oxygen atoms. The gas-phase species H, D, O, and CO are singled out for investigation; these species are allowed to accrete onto dust grains with rates defined in Caselli et al. (1998), while the accretion of other gas-phase species is not considered at all. Despite their accretion onto dust grains, H, D, O, and CO maintain constant gas-phase abundances. This apparently contradictory assumption is legitimate as long as the duration of the calculation does not exceed the 1/e depletion time scale of $10^9/n(H_2)$ yr, so that any diminution of gas-phase abundances is not significant. The assumption is made to simplify the mathematics; once depletion is considered, gas-phase chemical processes to reform the gas-phase species must also be included.

Three sets of gaseous abundances are considered; these are taken from steady-state cloud models of the gas phase (Lee et al., 1996) and are labeled the low, intermediate, and high density cases. The abundances are listed in Table 1.

Table 1. Gas-phase abundances ($cm^{-3}$)

| Species | Low density | Intermediate density | High density |
|---|---|---|---|
| $n_H = n(H) + 2n(H_2)$ | $10^3$ | $10^4$ | $10^5$ |
| $n(H)$ | 1.15 | 1.15 | 1.10 |
| $n(O)$ | 0.09 | 0.75 | 7.0 |
| $n(CO)$ | 0.075 | 0.75 | 7.5 |

The abundance of D in the gas has been varied from a concentration $n(D) = 0.01$ $cm^{-3}$ to 1.0 $cm^{-3}$; such high abundances of atomic deuterium can be achieved by gas-phase fractionation, especially at high density when the $DCO^+/HCO^+$ abundance ratio becomes quite large (Roberts & Millar, 2000). Over very long periods of time, of course, gas-phase abundances also change due to reactions and accretion onto grains, and it is necessary to consider complete models in which both gas-phase and surface chemistry occur (Ruffle & Herbst, 2000). At high density, for example, the depletion time is ≈ 2 x $10^4$ yr, so that a restricted model of the sort considered here is not useful for times longer than this.

Upon accretion onto the dust, the four gas-phase species H, O, D, and CO can evaporate or react to form the following stable molecules: $H_2$, HD, $D_2$, $H_2O$, HDO, $D_2O$, $CO_2$, $H_2CO$, HDCO, $D_2CO$, $CH_3OH$, $CH_3OD$, $CH_2DOH$, $CHD_2OH$, $CH_2DOD$, $CHD_2OD$, $CD_3OH$, $CD_3OD$, plus transient radicals such as HCO, DCO, $CH_3O$, etc. Evaporation of the produced species is not considered. The chemistry is assumed to be diffusive in nature and to occur on grains with radius 0.1μ and $10^6$ active binding sites (Hasegawa et al., 1992). Calculations are undertaken for dust temperatures in the range 10-20 K. The chemistry has been followed for times through $10^{3-5}$ yr, at which time significant mantles have developed. The diffusive reactions, which are almost all of the form A + B → C, are of two types: those that occur without activation energy and those that require it. The reactions are listed in Table 2, with estimated activation energies if appropriate. The reactions H + CO and H + $H_2CO$ have uncertain activation energies, labelled E1 and E2 respectively, in the range 1000-2000 K (Woon 2001); we have used 2000 K in our standard calculation. Reactions in which D atoms partially replace the H atoms have different activation energy barriers than E1 and E2 due to zero-point vibrations. Calculated for us by D. Woon (private communication), these differences are accurate, regardless of the accuracy of the barriers for the H + CO and H + $H_2CO$ reactions.

Table 2. Reactions in model

| Reactants | Products | Activation Energy (K) |
|---|---|---|
| H + CO | HCO | 1000-2000 (E1) |
| D + CO | DCO | E1 - 70 |
| H + HCO | $H_2CO$ | ---------- |
| D + HCO | HDCO | ---------- |
| H + DCO | HDCO | ---------- |
| D + DCO | $D_2CO$ | --------- |
| H + $H_2CO$ | $CH_3O$ | 1000-2000 (E2) |
| D + $H_2CO$ | $CH_2DO$ | E2 - 201 |
| H + HDCO | $CH_2DO$ | E2 - 35 |
| D + HDCO | $CHD_2O$ | E2 – 242 |
| H + $D_2CO$ | $CHD_2O$ | E2 – 75 |
| D + $D_2CO$ | $CD_3O$ | E2 – 287 |
| H + $CH_3O$ | $CH_3OH$ | ----------- |
| D + $CH_3O$ | $CH_3OD$ | ----------- |
| H + $CH_2DO$ | $CH_2DOH$ | ----------- |
| D + CH$_2$DO | CH$_2$DOD | ---------- |
| H + CHD$_2$O | CH$_2$DOH | ---------- |
| D + CHD$_2$O | CHD$_2$OD | ---------- |
| H + CD$_3$O | CD$_3$OH | ---------- |
| D + CD$_3$O | CD$_3$OD | ---------- |
| H + H | H$_2$ | ---------- |
| D + H | HD | ---------- |
| D + D | D$_2$ | ---------- |
| H + O | OH | ---------- |
| D + O | OD | ------------- |
| H + OH | H$_2$O | ----------- |
| D + OH | HDO | ----------- |
| H + OD | HDO | ---------- |
| D + OD | D$_2$O | ---------- |
| O + O | O$_2$ | ---------- |
| O + CO | CO$_2$ | 1000 |
| O + HCO | CO$_2$ + H | ---------- |
| O + DCO | CO$_2$ + D | ---------- |

Diffusive rate coefficients for each of the reactions are related to the sum of the rates of diffusion via tunneling and hopping over the entire grain for each reactant (Hasegawa et al., 1992; Caselli et al., 1998) multiplied by a factor $\kappa$ for tunneling under the chemical activation energy barrier if it exists (Tielens and Hagen, 1982). The diffusion rates for all species considered here but deuterium have been discussed previously; we start with the three sets of rates labeled "fast", "slow H", and "slow" by Ruffle and Herbst (2000). During the time periods considered here, both the "slow H" and "slow" rates do not allow the production of significant amounts of formaldehyde and methanol, although Ruffle and Herbst (2000) show that they can still be produced but over much longer time periods. We consequently report results here for fast rates only. We assume that the desorption energy and barrier against diffusion for D are both higher than that of H by the zero-point energy difference in the physisorption well, which is 21 K in the harmonic approximation when the surface vibrational frequency of Hasegawa et al. (1992) is utilized. The net result is that at 10 K, the fast diffusion, occurring by tunneling, and evaporation rates of D are approximately an order of magnitude lower than those of atomic hydrogen. For reactions with activation energy barriers, D atoms react considerably more slowly than their H counterparts due mainly to the lower tunneling efficiency of the heavier species. For example, for the reactions H + CO and D + CO, $\kappa$ is calculated to be 2 x $10^{-8}$ and 3 x $10^{-11}$, respectively, if E1 = 2000 K. These differences between D and H mostly affect doubly deuterated species because their formation requires one slow D reaction to proceed. Singly deuterated species such as HDCO, on the other hand, form efficiently via slow reactions involving H and reactions between D atoms and radicals without activation energy.

## 3. Methods of Calculation

### 3.1 *Modified Rate Approach*

This approach starts with the rate equation method (Pickles & Williams, 1977; Hasegawa et al., 1992; Caselli et al., 1998). Diffusive rate equations and rate laws are modified in a semi-empirical manner to reproduce the probabilistic nature of surface chemistry, which arises from the small number of reactive species present on a grain at any one time (Caselli et al., 1998; Stantcheva et al., 2001). The strength of the method lies in the ease of utilizing it in gas-grain chemical models; its weakness is that semi-empirical corrections to a complex series of equations can never be truly complete for all situations.

Although the modified rate approach to reactions without activation energy has been discussed in the literature, the approach to reactions with activation energy has not been fully discussed. The basic idea of the modified rate treatment of reactions with activation energy (e.g. H + CO) is to use the standard rate equation method until the weakly reactive species (e.g. CO) becomes so abundant that the probability of reaction exceeds unity. When this happens during the smaller of an accretion interval between landings of reactive species (H, D, O) and the evaporation time of the reactive species, the rate law is changed so that the maximum value of the reaction probability is one. An unintended and artificial consequence of this approach is an enhancement in the surface population of reactive species such as H, but since their abundances are small and unobservable anyway, the consequence is not an important one.

Let us consider the H-CO reaction on a grain that contains populations $N$ of both CO and $H_2CO$. When these populations are low enough that the probability of reaction with an accreting H is small, the normal rate law for HCO formation pertains; viz.,

$$\frac{dN(HCO)}{dt} = (k_{diff}(H) + k_{diff}(CO))\kappa_{H-CO}N(H)N(CO), \qquad (1)$$

where the diffusion rates $k_{diff}$ pertain to motion over the whole grain, $\kappa_{H-CO}$ is the probability of tunneling under the H-CO potential barrier (Tielens and Hagen, 1982), and the H diffusion rate far exceeds that of CO so that the latter can be neglected. In considering the modified rate law, let us first neglect evaporation of H and let us assume that the H-CO reaction is the first to achieve a probability of occurring that is greater than unity during an accretion interval, as determined by the normal rate method. When this happens, the rate law is modified to

$$\frac{dN(HCO)}{dt} = k'N(H) \times \frac{\kappa_{H-CO}N(CO)}{\kappa_{H-CO}N(CO) + \kappa_{H-H_2CO}N(H_2CO) + \kappa_{H-HDCO}N(HDCO) + \kappa_{H-D_2CO}N(D_2CO)} \qquad (2)$$

where $\kappa_{H-H2CO}$ stands for the tunneling probability under the H-$H_2CO$ potential barrier, $\kappa_{H-HDCO}$ and $\kappa_{H-D2CO}$ are similarly defined, and $k'$ is the total accretion rate of reactive species (s$^{-1}$). The fraction on the right-hand side of equation (2) yields the probability that H reacts with CO rather than with $H_2CO$, HDCO, or $D_2CO$. Note that if the H-CO reaction is the dominant one, then the fraction goes to unity and the rate of HCO formation is related to an overall accretion rate multiplied by the surface population of H atoms per grain. Note also that the CO population is no longer in the rate law. If reactive species other than H do not accrete onto a grain at an appreciably greater rate than H (a condition that is almost always met) and the H atoms can only react with CO, the rate of formation of HCO in equation (2) will indeed approximately equal the accretion rate of H atoms at steady state. To see this, consider the rate law for the surface population of hydrogenic atoms. These atoms are produced by accretion and depleted mainly by reaction with CO in the limit we are considering, so that

$$\frac{dN(H)}{dt} \approx k' - k'N(H) , \qquad (3)$$

if other loss mechanisms are not considered. At steady state, when $\frac{dN(H)}{dt} = 0$, the surface population of H is near unity and equation (2) reduces to the accretion rate of hydrogen atoms. Once the H-CO rate is treated by a modified rate law, then the reactions $H-H_2CO$, H-HDCO, and $H-D_2CO$ must also be so treated. A similar analysis holds for reactions with barriers involving D and CO, $H_2CO$, etc.

Now let us consider the possibility of evaporation of reactant H. If the evaporation rate of H is larger than the accretion rate of reactive species, then the diffusion rate of H must be reduced only to the evaporation rate $k_{evap}$ of this atom (Caselli et al., 1998) rather than to the accretion rate. In equations (2) and (3), $k'N(H)$ must now be replaced by $k_{evap}N(H)$ and there is an additional loss term in (3) due to the actual evaporation of H atoms. (Other reactive loss mechanisms are not considered.) The steady-state formation rate of HCO is then approximately 1/2 of the accretion rate of H, which is 1/2 of the desired result. A more appropriate reduction of the diffusion rate of H would be to a value somewhat greater than the evaporation rate of this species, but in practice the reduction chosen does not result in serious error.

The discussion above follows and extends the modified rate treatment of Stantcheva et al. (2001). The earlier treatment of Caselli et al. (1998; unpublished work) has some additional semi-empirical factors. In general, we find little difference between the two approaches; where small differences exist, we report the values obtained from the method of Stantcheva et al.

### 3.2 *Monte Carlo Approach*

Up to the present, we have utilized a pseudo-Monte Carlo approach. Based on the work of Tielens & Hagen (1982), this approach follows individual species as they land on a grain and computes the relative probabilities that this species will undergo specific reactions, evaporate, or do nothing before another species lands (Stantcheva et al. 2001). It is applicable only in the accretion limit, where there is on average less than one reactive species per grain, because it is does not follow reactions between two species that linger on the grain for more than one accretion period. For this paper, we have switched to a more general Monte Carlo approach (Gillespie, 1976), in general even more computer-intensive, based on the state of the system and the mean field approximation. This approach is similar to that utilized by Charnley (1998) for gas-phase chemistry. The details of the procedure will be introduced in a future paper along with our direct solution of the master equation (Shematovich et al., in preparation).

## 4. Results

We refer to the following physical conditions as our standard model: an elapsed time of $10^4$ yr, high density, E1=E2=2000 K, a dust temperature of 10 K, and $n(D) = 0.3$ cm$^{-3}$. For all conditions studied at times greater than $10^3$ yr, steady-state conditions pertain (atoms and radicals at constant surface population, major species increasing linearly with time and at the same rate). In Table 3, we list the populations of assorted non-deuterated molecules, in monolayers, or $10^6$ particles, per grain, achieved at $10^4$ yr for the standard case as well as intermediate and low densities with both theoretical approaches. It can be seen that, except for the high-density case, essentially all of the CO landing on grains is converted into methanol. This result occurs because the accretion rate of H atoms exceeds that of CO molecules. At high density, on the other hand, there is a competition among CO, $H_2CO$, and methanol, the populations of which are in good agreement with the ratios in Figure 2 of Charnley et al. (1997). The results in Table 3 show that the modified rate values do not always agree precisely with the Monte Carlo values, although the worst disagreements are a factor of 2-3. The total abundances shown in Table 3 are larger than the sums of the individual abundances listed because of surface molecular hydrogen, HD and other deuterated species not contained in the table. Note that water ice is not the dominant constituent at high density in the outer mantle layers considered here; most of the water has already been formed at earlier stages when the H atom density in the gas is greater.

Table 3. Population in monolayers achieved by $10^4$ yr at 10 K

| Species | High Density Monte Carlo | High Density Modified rate | Intermediate Density Monte Carlo | Intermediate Density Modified rate | Low Density Monte Carlo | Low Density Modified rate |
|---|---|---|---|---|---|---|
| CO | 5.0 | 4.4 | 0.0 | 0.0 | 0.0 | 0.0 |
| $H_2O$ | 1.4 | 0.75 | 0.45 | 0.30 | 0.070 | 0.070 |
| $O_2$ | 2.7 | 2.7 | 0.080 | 0.13 | 0.0013 | 0.0019 |
| $CO_2$ | 0.67 | 1.3 | 0.055 | 0.15 | 0.00075 | 0.0013 |
| $H_2CO$ | 0.44 | 0.27 | 0.0 | 0.0 | 0.0 | 0.0 |
| $CH_3OH$ | 0.088 | 0.089 | 0.42 | 0.34 | 0.046 | 0.046 |
| Total | 11.1 | 10.1 | 1.45 | 1.28 | 0.31 | 0.29 |

In Table 4, we compare the abundance ratios (designated $f$) between deuterated and normal isotopomers for the same model parameters. Values close to zero or undefined are designated by dashes, as are small values for which the Monte Carlo method has poor statistics. Here the agreement between the two theoretical treatments is excellent. At the large abundance chosen for $n(D)$, the values of $f$ are very high for high gas density; while for lower densities, the $f$ values for the singly deuterated isotopomers remain high while those for the doubly and multiply deuterated isotopomers of methanol become much lower. Of course, high values for $n(D)$ are only associated with high density because it is necessary for significant depletions of heavy materials from the gas phase to occur (Roberts and Millar, 2000). So, in actuality, large $f$ values cannot be expected except at high densities unless we do not understand the conditions under which large values of $n(D)$ occur. Because the calculated $f$ values for $CHD_2OH$ and $CH_2DOD$ both exceed that for $D_2CO$ at high density, these species are definitely worth searching for, especially since normal methanol can be quite abundant in warm sources (Caselli et al., 1993; van Dishoeck et al., 1995; van der Tak et al. 2000). It is even possible that the two triply deuterated isotopomers can be detected!

Table 4. Abundance ratios $f$ for deuterated species at $10^4$ yr and 10 K ($n(D) = 0.3$ cm$^{-3}$) compared with normal isotopomers

| Species | High Density Monte Carlo | High Density Modified rate | Intermediate Density Monte Carlo | Intermediate Density Modified rate | Low Density Monte Carlo | Low Density Modified rate |
|---|---|---|---|---|---|---|
| $f$HDCO | 0.33 | 0.32 | ----------- | ------------ | ---------- | ---------- |
| $f$D$_2$CO | 0.026 | 0.025 | ----------- | ------------ | ---------- | ---------- |
| $f$CH$_3$OD | 0.20 | 0.18 | 0.19 | 0.22 | 0.19 | 0.17 |
| $f$CH$_2$DOH | 0.70 | 0.74 | 0.19 | 0.22 | 0.18 | 0.17 |
| $f$CHD$_2$OH | 0.17 | 0.18 | 3.3(-04) | 2.1(-04) | ---------- | 1.3(-04) |
| $f$CH$_2$DOD | 0.14 | 0.13 | 0.034 | 0.048 | 0.034 | 0.030 |
| $f$CHD$_2$OD | 0.034 | 0.033 | 6.6(-05) | 4.5(-05) | 2.2(-05) | 2.3(-05) |
| $f$CD$_3$OH | 0.014 | 0.015 | ----------- | 4.5(-08) | ---------- | 2.3(-08) |
| $f$CD$_3$OD | 0.0025 | 0.0028 | ----------- | 1.0(-08) | ---------- | 4.0(-09) |
| $f$HDO | 0.39 | 0.40 | 0.38 | 0.46 | 0.39 | 0.37 |
| $f$D$_2$O | 0.038 | 0.039 | 0.037 | 0.053 | 0.039 | 0.033 |

In Table 5, we list the populations for the stable non-deuterated molecules and the $f$ values for the deuterated isotopomers computed with the modified rate method when the activation energies E1 and E2 are varied from the standard values of 2000 K and the other standard conditions are maintained. Differences are minimal when E1 and E2 are both lowered to 1000 K. When E1 is lowered to 1000 K and E2 is held constant at 2000 K, the conversion of surface formaldehyde into methanol becomes inefficient and the $f$ values of some but not all deuterated methanol isotopomers decrease sharply. The reason for the inefficient formation of methanol can be deduced from equation (2); when the barrier for hydrogenating formaldehyde exceeds that for hydrogenating CO, then most H atoms react with CO and not with formaldehyde. Analogous results for the $f$ values of HDCO, $CH_3OD$, $CH_2DOH$, and $D_2CO$ from a calculation in which E2 ($\approx$650K) is less than E1 were presented in Figure 3 of Charnley et al. (1997) for high density conditions. The agreement with our E1=E2=2000K results is reasonable except for the case of $D_2CO$, where our value for $f$ is larger by a factor of a few. Further comparison is discussed below for a range of $n(D)$.

Table 5. Effects of changes in activation energies E1, E2 (modified rates, high density, 10 K, $n(D) = 0.3$ cm$^{-3}$) on populations (monolayers) and abundance ratios of deuterated isotopomers

| Species | 2000K, 2000K | 1000 K, 1000 K | 1000 K, 2000 K |
|---|---|---|---|
| Populations | | | |
| CO | 4.4 | 4.4 | 4.1 |
| $H_2O$ | 0.75 | 0.75 | 0.76 |
| $O_2$ | 2.7 | 2.7 | 2.6 |
| $CO_2$ | 1.3 | 1.3 | 1.4 |
| $H_2CO$ | 0.27 | 0.26 | 0.49 |
| $CH_3OH$ | 0.089 | 0.087 | 0.0011 |
| Total | 10.1 | 10.1 | 9.94 |
| Abundance Ratios | | | |
| $f$HDCO | 0.32 | 0.29 | 0.37 |
| $fD_2CO$ | 0.025 | 0.020 | 0.033 |
| $fCH_3OD$ | 0.18 | 0.18 | 0.18 |
| $fCH_2DOH$ | 0.74 | 0.86 | 0.47 |
| $fCHD_2OH$ | 0.18 | 0.25 | 0.067 |
| $fCH_2DOD$ | 0.13 | 0.15 | 0.15 |
| $fCHD_2OD$ | 0.033 | 0.043 | 0.043 |
| $fCD_3OH$ | 0.015 | 0.023 | 0.0025 |
| $fCD_3OD$ | 0.003 | 0.004 | 0.0005 |
| $f$HDO | 0.40 | 0.39 | 0.41 |
| $fD_2O$ | 0.039 | 0.038 | 0.041 |

Figures 1 and 2 show what happens to the populations of assorted species when the temperature of our standard model is varied in the range 10-20 K. In Figure 1, we plot the mole fractions (relative populations) of the major species rather than their populations since the total population does not change greatly with temperature, while in Figure 2, we plot the $f$ values of deuterated isotopomers. Both the Monte Carlo (Figure 1a and 2a) and modified rate approaches (Figures 1b and 2b) have been utilized. From a comparison of Figures 1a and 1b, it can be seen that the agreement between the two approaches for the non-deuterated species is only uniformly good (factor of $\approx 2$) for temperatures in the range 10-15 K. At higher temperatures, the species with rapidly declining abundances ($CO_2$, $H_2CO$, and $CH_3OH$) start their rapid declines at lower temperatures (the offset is $\approx$ 2 K) in the Monte Carlo approach. Agreement for the major species (CO, $O_2$, $H_2O$) remains good. From a comparison of Figures 2a and 2b, we note that the two methods agree satisfactorily for most isotopomers throughout the whole temperature range (statistics become too poor for the Monte Carlo method to be viable for some species at higher temperatures). Exceptions occur for two doubly deuterated isotopomers plotted ($D_2CO$ and $CHD_2OH$); although the agreement is good at 10 K, the Monte Carlo results start to decline considerably at lower temperature than do the modified rate values. Once again the offset is about 2 K. This effect occurs for the other doubly and multiply deuterated isotopomers of methanol as well. Here the use of the $f$ value exacerbates the discrepancy because the modified rate and Monte Carlo results tend to differ in the opposite sense for the normal and deuterated isotopomers.

Ignoring the differences between the two techniques, we can state qualitatively that among the major species, it becomes relatively more difficult to produce formaldehyde and methanol above 15-17 K in the time considered because of the faster evaporation of atomic hydrogen. The $f$ values show a complex pattern; most singly deuterated species do not exhibit a strong dependence with temperature while most doubly deuterated species show a very strong inverse dependence with an increase in temperature. Presumably the explanation for this dichotomy lies in the fact that production of the doubly deuterated species requires at least one reaction with activation energy involving atomic D whereas the production of the singly deuterated species can occur without any such reaction. The distinction is critical because the reaction probability for slow reactions with D is much lower than that for slow reactions with H, a more efficient tunneler, and this lower efficiency competes less favorably with evaporation as temperature increases.

Figure 3 shows the dependence of our standard-model results for assorted $f$ values of singly and doubly deuterated isotopomers on the concentration of deuterium atoms in the range 0.01-1.0 cm$^{-3}$. Since the Monte Carlo and modified rate methods are in agreement here, we have only reported results from the latter. As first reported by Charnley et al. (1997), it can be seen that the $f$ values for the doubly deuterated isotopomers rise at a slope double that of the singly deuterated isotopomers in the log-log plot. Thus, at very high atomic D/H ratios, the $f$ values for the doubly deuterated species approach those of their singly deuterated counterparts. Assuming that the standard conditions we use can be applied to the dense cold gas in the Orion compact ridge and IRAS 16293-2422 prior to grain accretion, the gas-phase concentrations of atomic D needed to reproduce the observed $f$ values for D$_2$CO are 0.1 cm$^{-3}$ and 0.6 cm$^{-3}$, respectively, which are roughly half of the values obtained by Charnley et al. (1977). This simple analysis contains the assumption that the $f$ values remain unchanged following evaporation of the surface species into the gas.

## 5. Conclusions

We have used the modified rate and Monte Carlo approaches to calculate how rapidly surface chemistry can produce large abundances of deuterated isotopomers of formaldehyde and methanol, especially at high gas densities ($n \approx 10^5$ cm$^{-3}$) and low temperatures (10 K). Indeed, within $10^4$ yr, surface chemistry can convert a high atomic D/H ratio in the gas to high abundance ratios for singly, doubly, and even multiply deuterated isotopomers. A search for doubly deuterated species not yet detected, especially those of methanol, thus seems warranted. Although the system of reactions leading to the deuterated formaldehyde and methanol isotopomers has been studied previously by Charnley et al. (1997), we have both extended and added to our understanding of this system and have used it as a test-bed for the semi-empirical modified rate method of diffusive surface chemistry. This approach, which so far has been the only method efficient enough to be used in large-scale models of time-dependent gas-phase and surface chemistry, works reasonably well when compared with a non-empirical Monte Carlo method. The differences are most apparent for rapid diffusion rates and high gas densities, the parameters and conditions emphasized here, where the H to CO abundance ratio in the gas and on grains is lowest. Even under these conditions, there are no major discrepancies for major surface species in the 10-20 K temperature range. Above 15 K, serious differences appear for species with declining abundances. Regarding *f* values, the only significant discrepancies concern minor species (doubly and multiply deuterated isotopomers at temperatures above 10 K). In general, although some differences obviously occur, especially at temperatures above 10 K, the serious problems associated with the normal rate method for even major species (Stantcheva et al., 2001) are eliminated. The level of agreement between the modified rate and Monte Carlo approaches for the deuterated methanol system under the most testing of conditions, though not uniform, lends some confidence that our semi-empirical approach is accurate for large gas-grain chemical models throughout much of parameter space.

To relate models of deuteration on grain surfaces to gas-phase observations in protostellar sources requires a knowledge of what happens in the gas phase after evaporation from the grain mantles. Our current understanding (Caselli et al., 1993) is that in sources such as the Orion Hot Core/compact ridge, associated with high-mass star formation, the gas-phase abundances of methanol and formaldehyde remain large for perhaps $10^{4-5}$ yr. Charnley et al. (1997) have modeled the gas-phase chemistry in the compact ridge, including the deuterated isotopomers; they find that through $10^5$ yr, changes in the *f* values from their surface values are limited to factors of a few. Although modeling as detailed as this has not yet been done for the protostellar source IRAS 16923-22422, around which the abundance of $D_2CO$ is known to be huge, Ceccarelli et al. (2001) have argued that "the present-day (gas-phase) $H_2CO$ and $D_2CO$ abundances very probably reflect the mantle composition."

Our results for surface methanol show that, at high gas density, this species is not as abundant as formaldehyde. Nevertheless, we calculate that in $10^4$ yr, 0.09 monolayers of methanol can be produced per grain, corresponding to a fractional abundance with respect to the total gas density of 1 x $10^{-7}$ with the normal grain parameters. This would seem ample for detection of methanol, its singly deuterated isotopomers, and even its doubly deuterated isotopomers, if the grain hypothesis is correct and our understanding of the gas-phase physical conditions prior to accretion onto the grains is not hopelessly inadequate. Methanol itself is highly abundant in hot core-type sources such as the Orion compact ridge (Menten et al., 1986; 1988), as well as in massive star forming regions (van der Tak et al., 2000). Moreover, it has been detected in IRAS 16293-2422 (van Dishoeck et al., 1995).

## Acknowledgments

We thank David Woon for calculating zero-point vibrational energies to enable us to estimate differences in activation energy barriers among isotopomers. P. C. acknowledges support from ASI (grants 98-116 and ARS-78-1999) and from MURST (project "Dust and Molecules in Astrophysical Environments.") E. H. would like to acknowledge the support of the National Science Foundation (US) for support of his program in astrochemistry, including the travel expenses of V. I. S. to visit the OSU laboratory. V. I. S. acknowledges support from the grant RFBR 01-02-16206.

**Figure Captions**

Figure 1a. Mole fractions of assorted major surface species calculated with the Monte Carlo method plotted as a function of temperature for standard (high density) conditions.

Figure 1b. Same as for Figure 1a except that the modified rate approach is used.

Figure 2a. Abundance ratios $f$ between deuterated and non-deuterated isotopomers calculated with the Monte Carlo method plotted as a function of temperature for standard conditions (high density with $n(D) = 0.03$).

Figure 2b. Same as for Figure 2a except that the modified rate approach is used.

Figure 3. Abundance ratios $f$ between deuterated and non-deuterated isotopomers plotted as a function of $n(D)$ for high density conditions at 10 K. The modified rate approach is used.

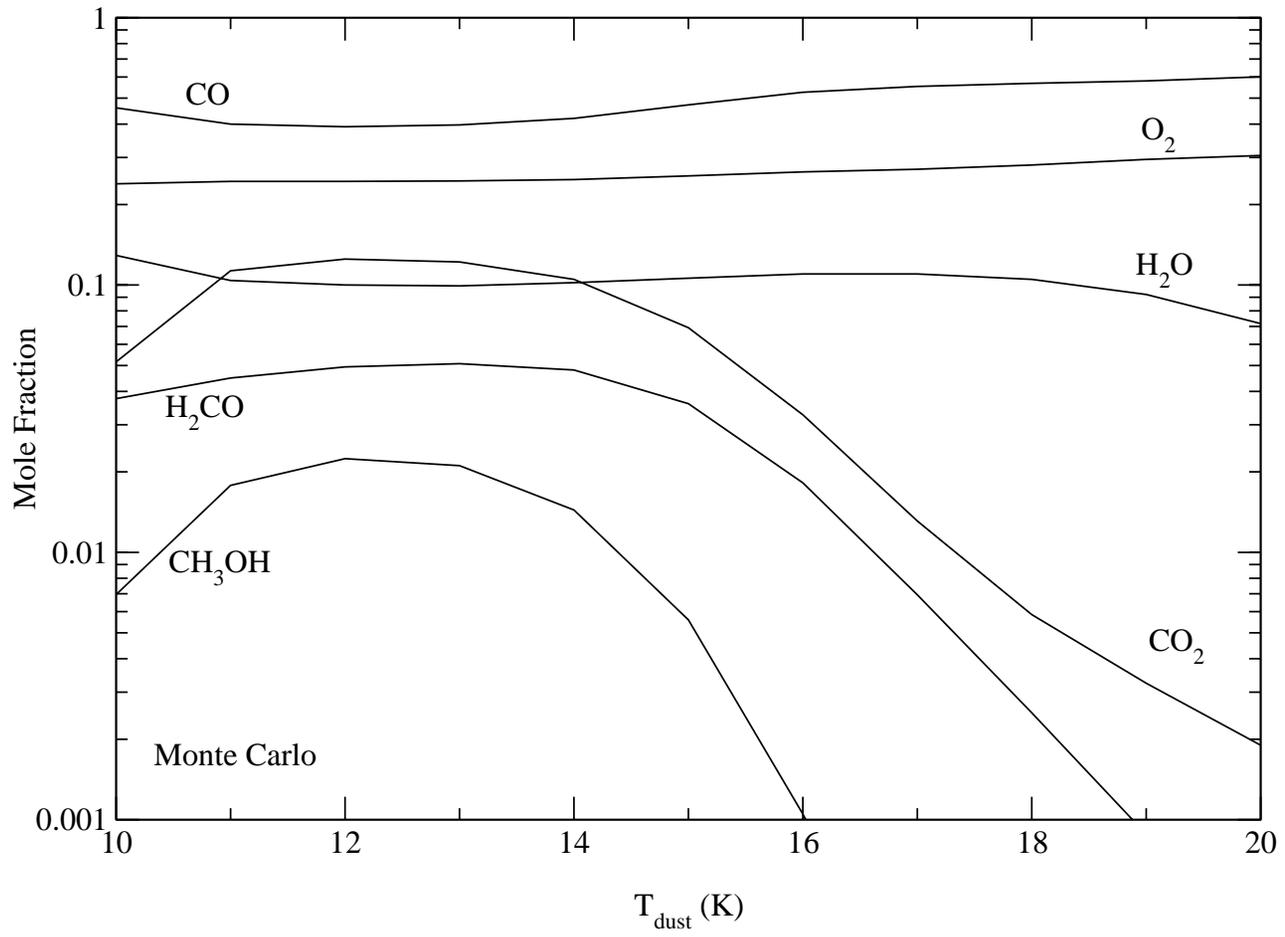

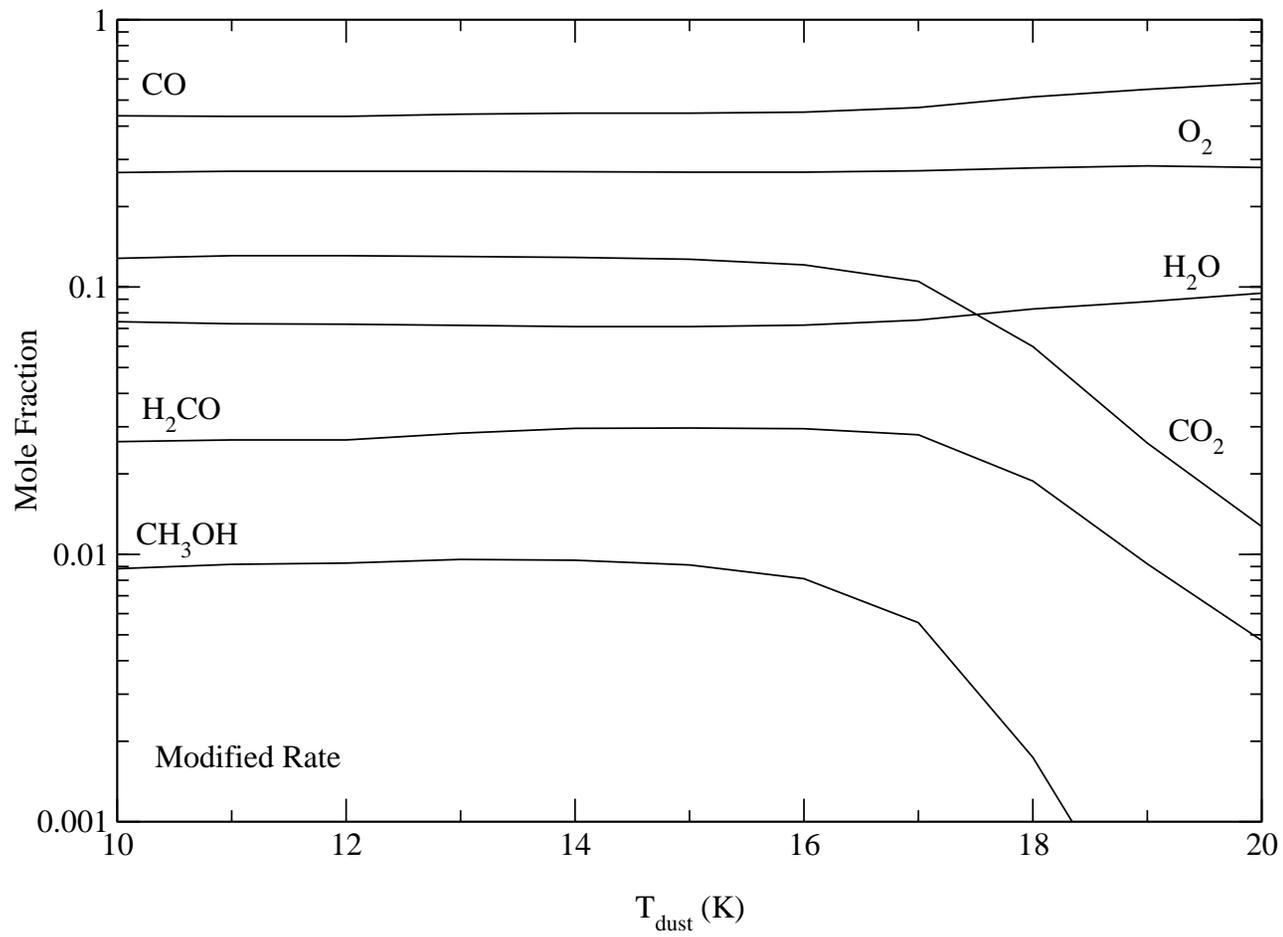
24

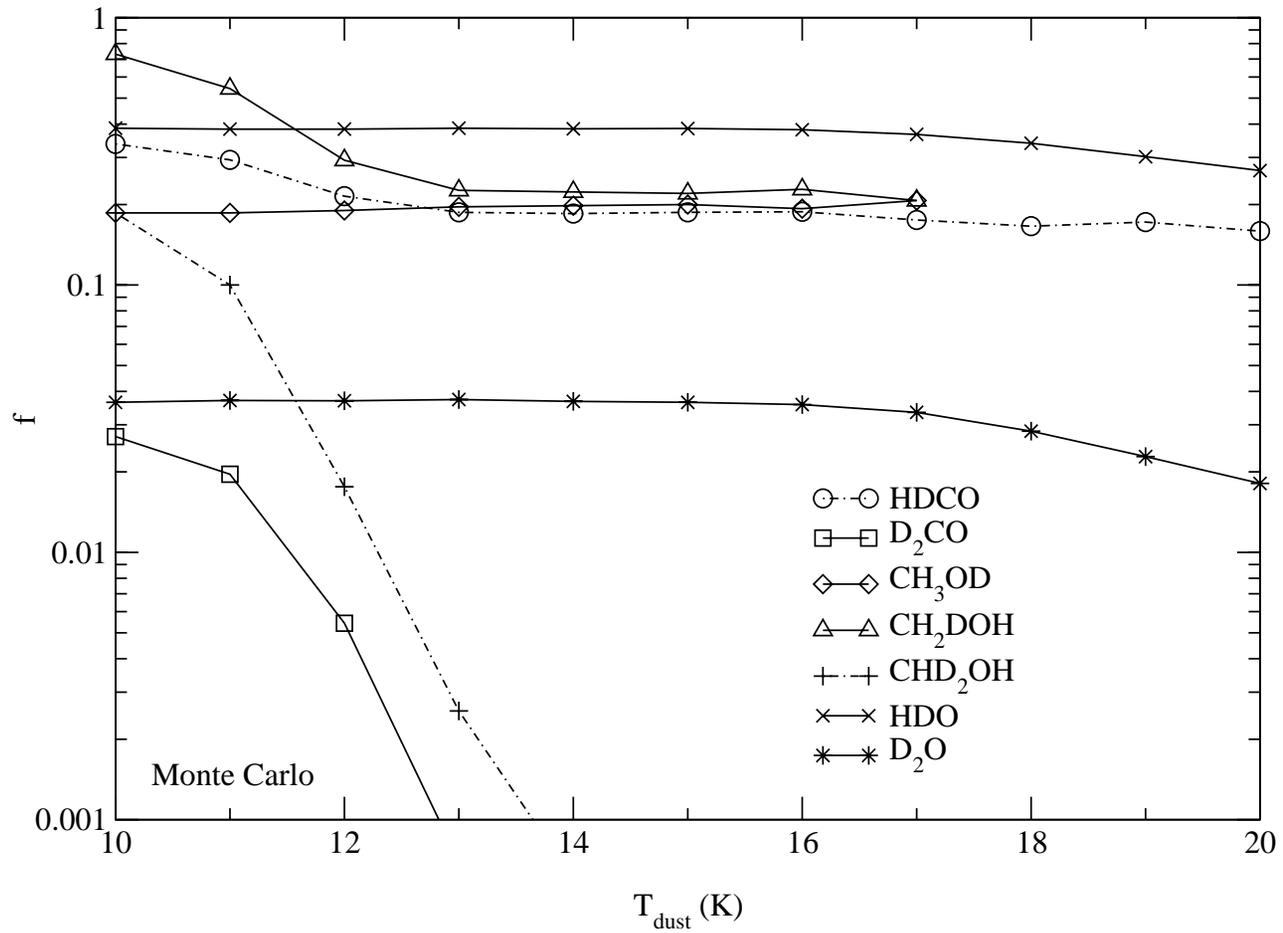

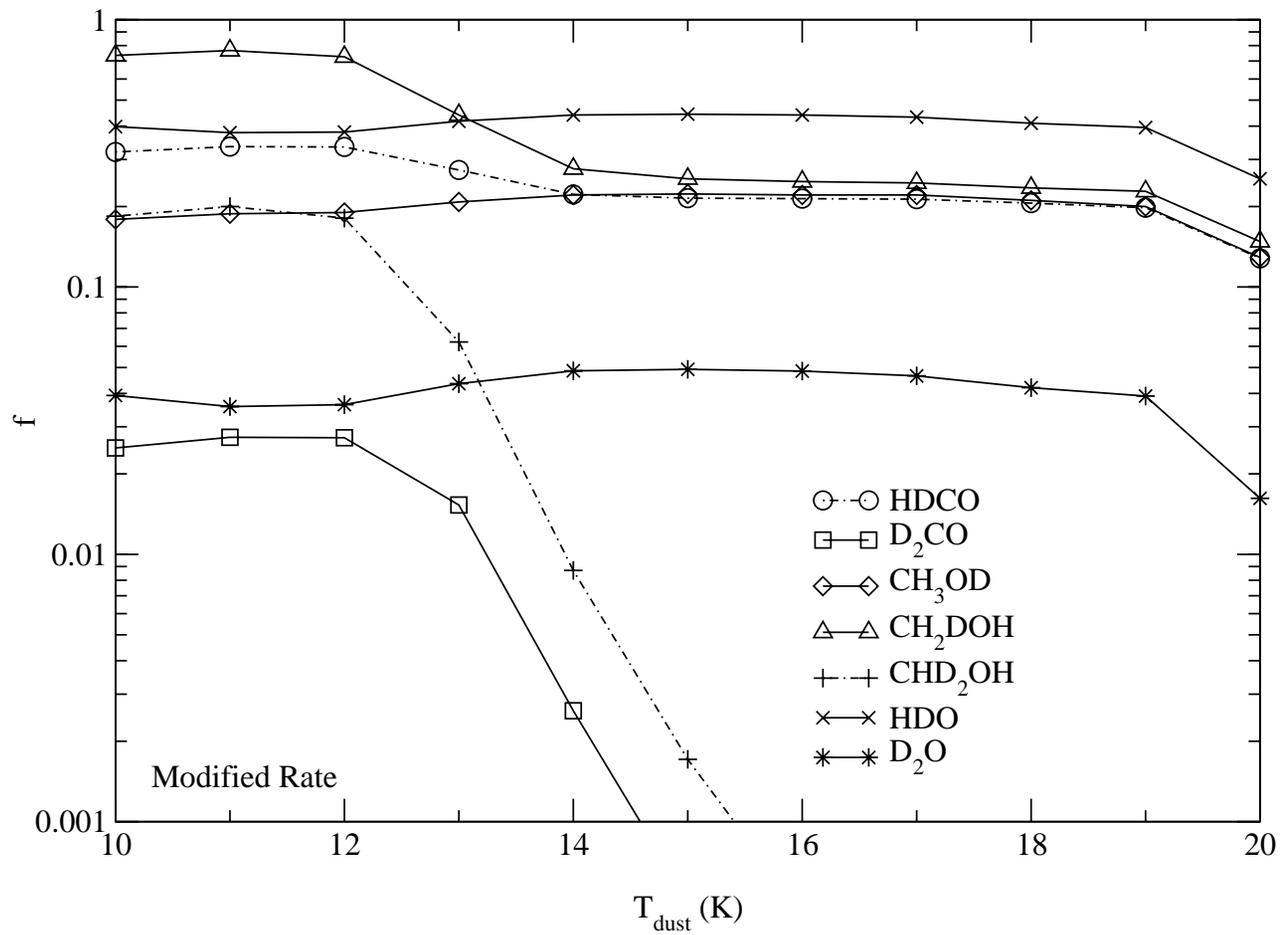

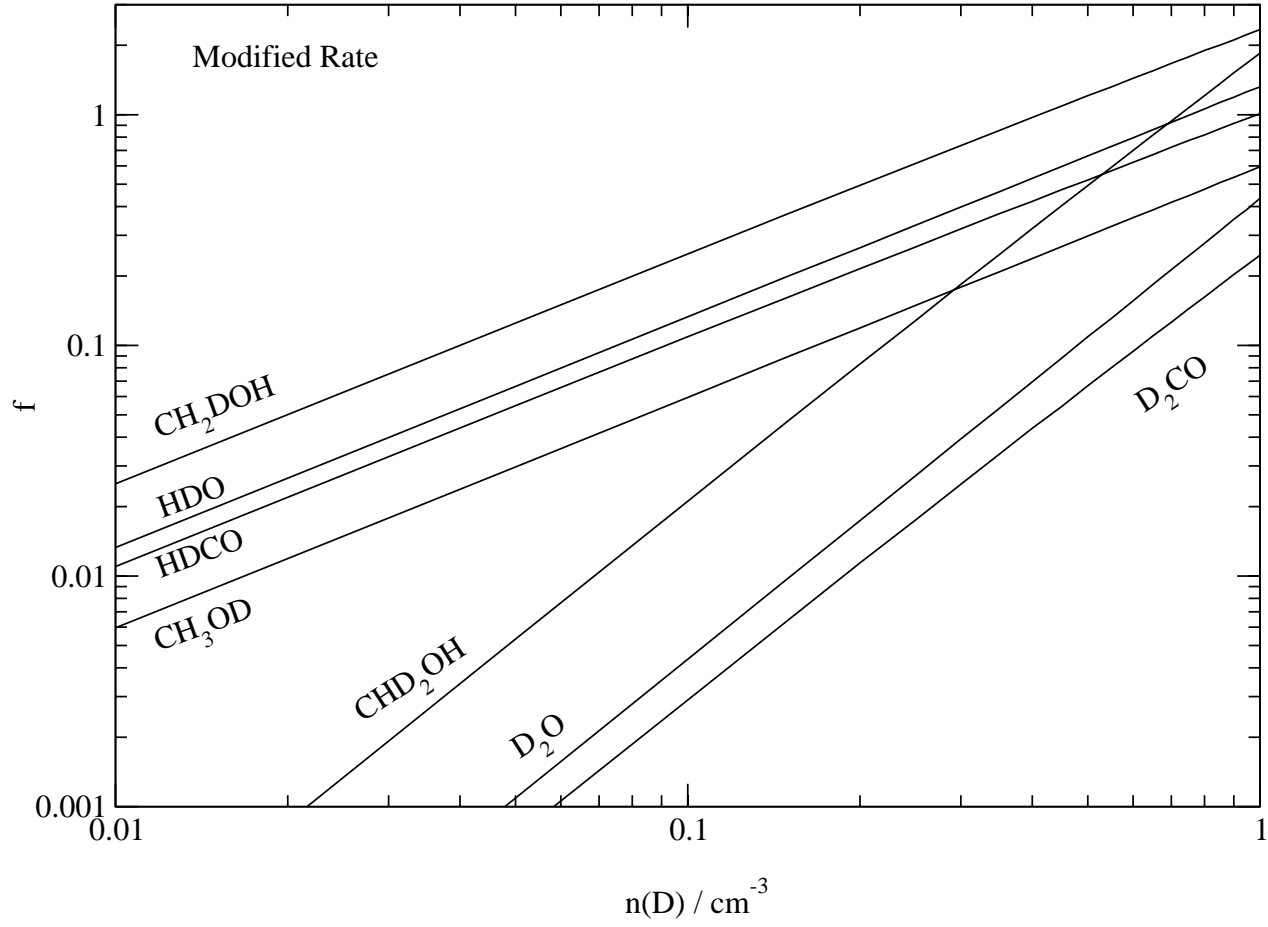